\documentclass[aps,prl,amsmath,amssymb,floatfix,twocolumn,amsmath,superscriptaddress,twocolumn,nofootinbib,tighten,letterpaper]{revtex4}
\usepackage{multirow}
\usepackage{bbold}
\usepackage{subfigure}
\usepackage{color}
\usepackage{mathrsfs}
\usepackage{hyperref}
\usepackage[normalem]{ulem}
\usepackage{bm}
\usepackage{pifont}

\usepackage{amsfonts, relsize, color}
\usepackage{graphics}
\usepackage{graphicx}
\usepackage{subfigure}
\usepackage{hyperref}
\usepackage{color}

\newcommand{\cmark}{\ding{51}}%
\newcommand{\xmark}{\ding{55}}%

\begin{document}
\title{\bf Higher-order topological superconductors in ${\mathcal P}$-, ${\mathcal T}$-odd quadrupolar Dirac materials}

\author{Bitan Roy}\email{bitan.roy@lehigh.edu}
\affiliation{Department of Physics, Lehigh University, Bethlehem, Pennsylvania, 18015, USA}

\date{\today}
\begin{abstract}
The presence or absence of certain symmetries in the normal state (NS) also determines the symmetry of the Cooper pairs. Here we show that parity (${\mathcal P}$) and time-reversal (${\mathcal T}$) odd Dirac insulators (trivial or topological) or metals, sustain a local or intra-unit cell pairing that supports corner (in $d=2$) or hinge (in $d=3$) modes of Majorana fermions and stands as a higher-order topological superconductor (HOTSC), when the NS additionally breaks discrete four-fold ($C_4$) symmetry. Although these outcomes does not rely on the existence of a Fermi surface, around it (when the system is doped) the HOTSC takes the form of a mixed parity, ${\mathcal T}$-odd (due to the lack of ${\mathcal P}$ and ${\mathcal T}$ in the NS, respectively) $p+id$ pairing, where the $p$($d$)-wave component stems from the Dirac nature of quasiparticles (lack of $C_4$ symmetry) in the NS. Thus, when strained, magnetically ordered Dirac materials, such as doped magnetic topological insulators (MnBi$_2$Te$_4$), can harbor HOTSCs, while the absence of an external strain should be conducive for the axionic $p+is$ pairing.     
\end{abstract}

\maketitle

\emph{Introduction}. The symmetry of the normal state (NS) plays an important role in understanding the nature of the Cooper wavefunctions inside the paired states at low temperatures~\cite{sigrist-ueda-rmp}. For example, (1) the lack of the inversion symmetry in the NS mixes the even- and odd-parity pairings~\cite{gorkov-rashba}, and (2) Dirac materials composed of linearly dispersing massive or massless quasiparticles in the NS, described by the angular momentum $\ell=1$ harmonics, harbor a plethora of exotic $p$-wave pairings~\cite{roy-ghorashi-foster-nevidomskyy,roy-alavirad-sau}, analogous to the ones in the B- and polar phases of $^3$He~\cite{volovik-book}. Here we investigate the role of time-reversal ($\mathcal T$), inversion ($\mathcal P$), as well as discrete crystalline four-fold ($C_4$) symmetry breaking on the paired states in two- and three-dimensional Dirac materials (insulator or metal) and come to the following conclusions.

\begin{figure*}[t!]
\subfigure[]{\includegraphics[width=0.24\linewidth]{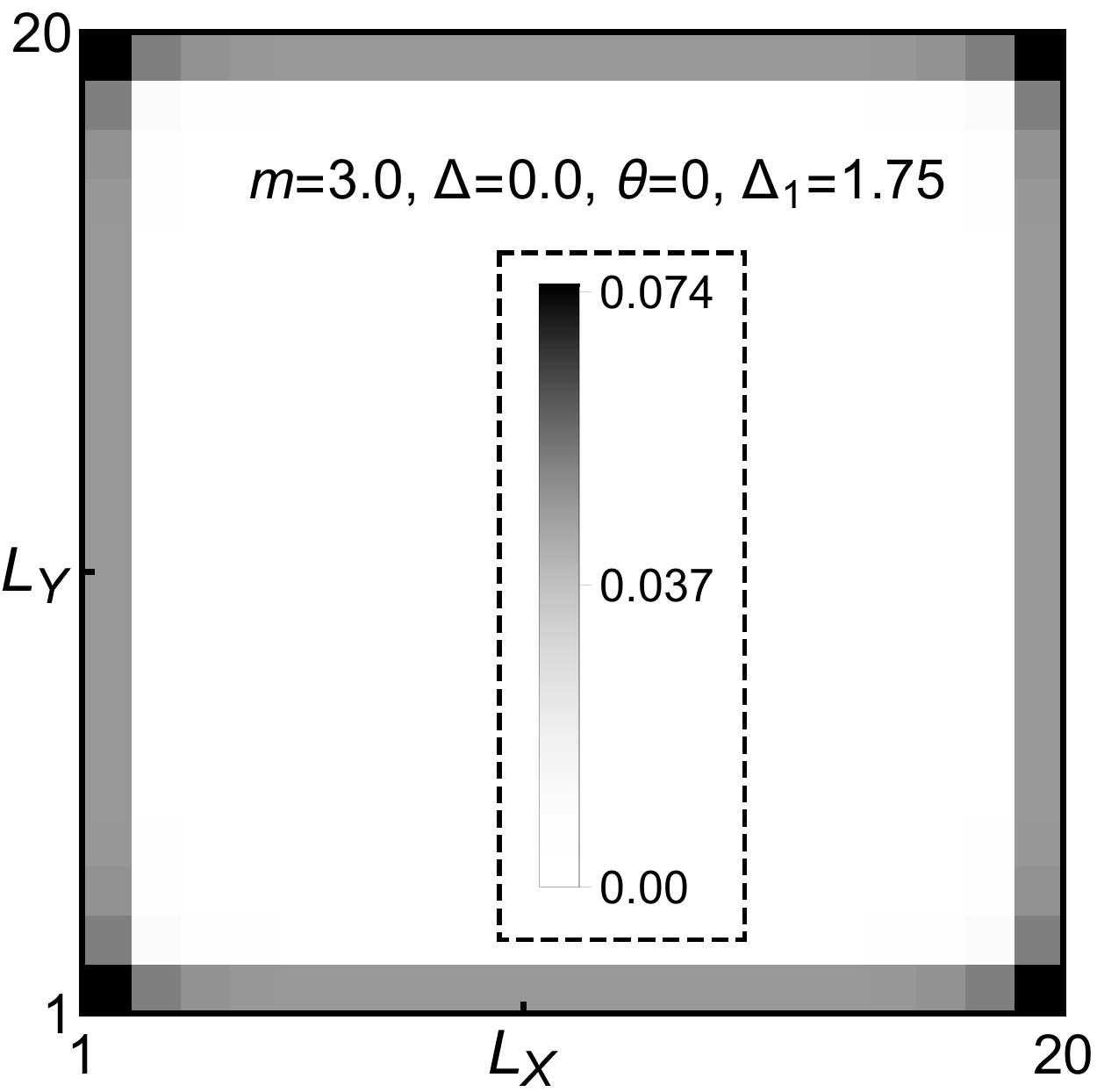}~\label{subfig1:delta1noDelta}}%
\subfigure[]{\includegraphics[width=0.24\linewidth]{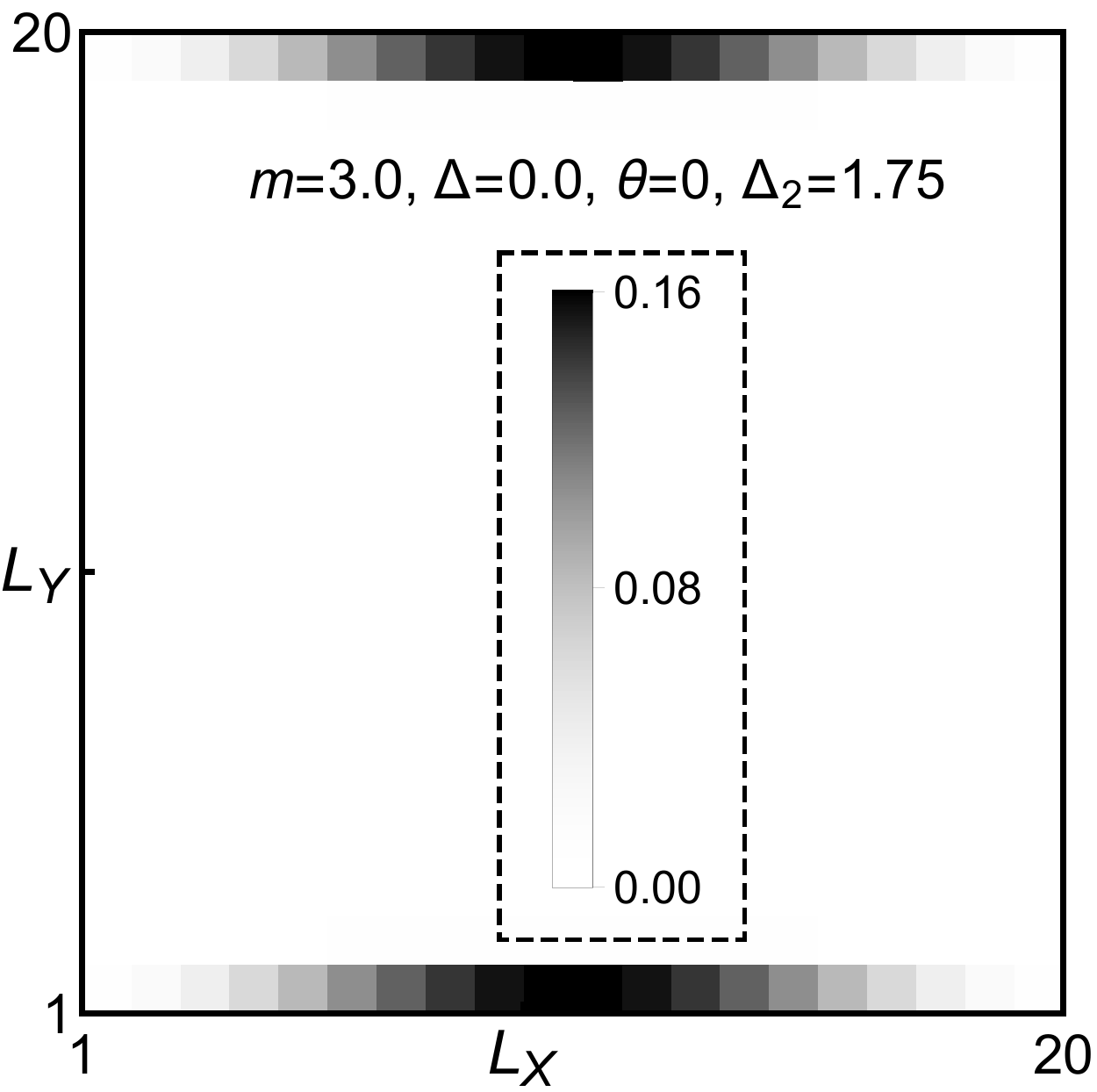}~\label{subfig1:delta2noDelta}}%
\subfigure[]{\includegraphics[width=0.24\linewidth]{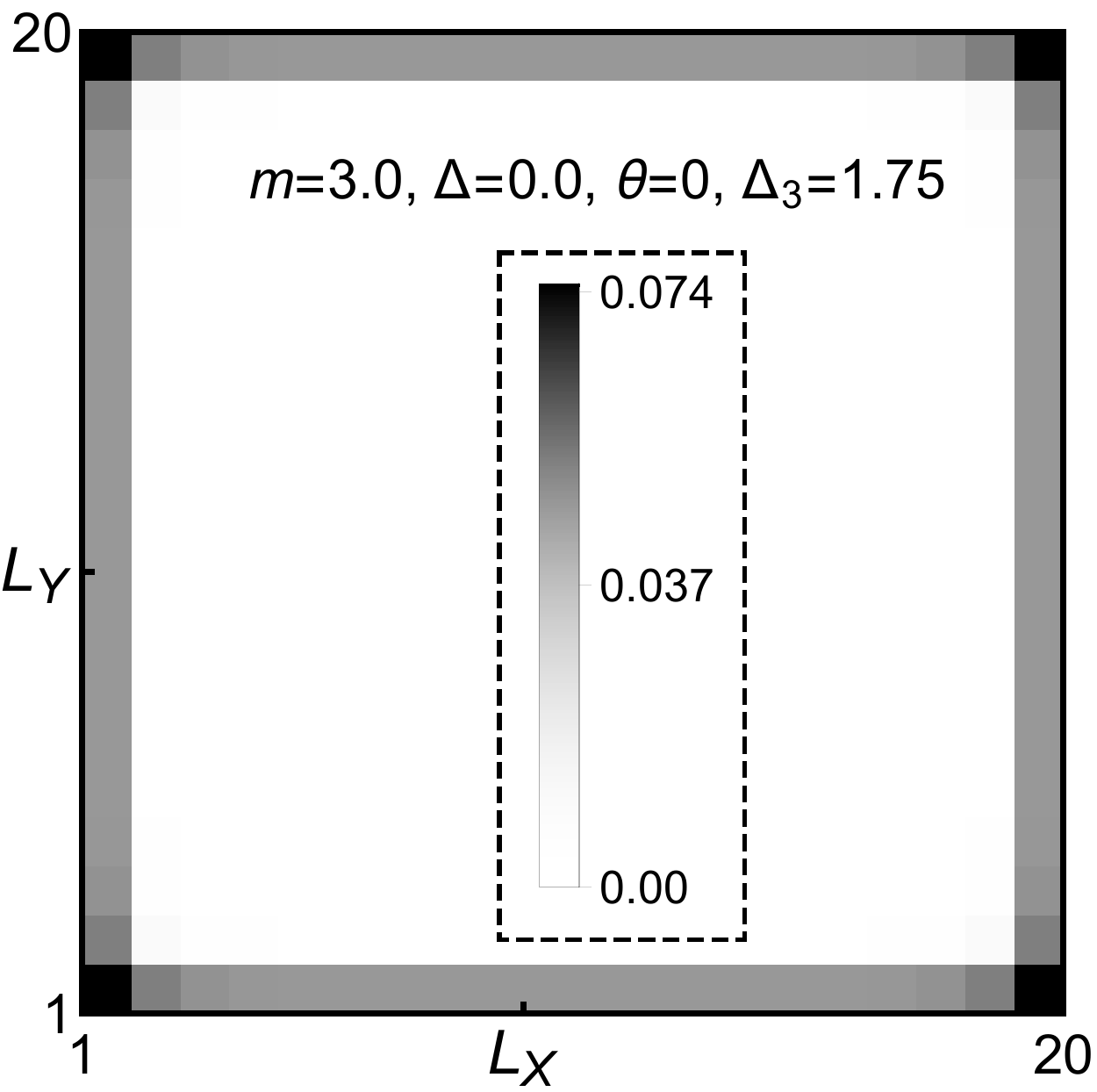}~\label{subfig1:delta3noDelta}}%
\subfigure[]{\includegraphics[width=0.24\linewidth]{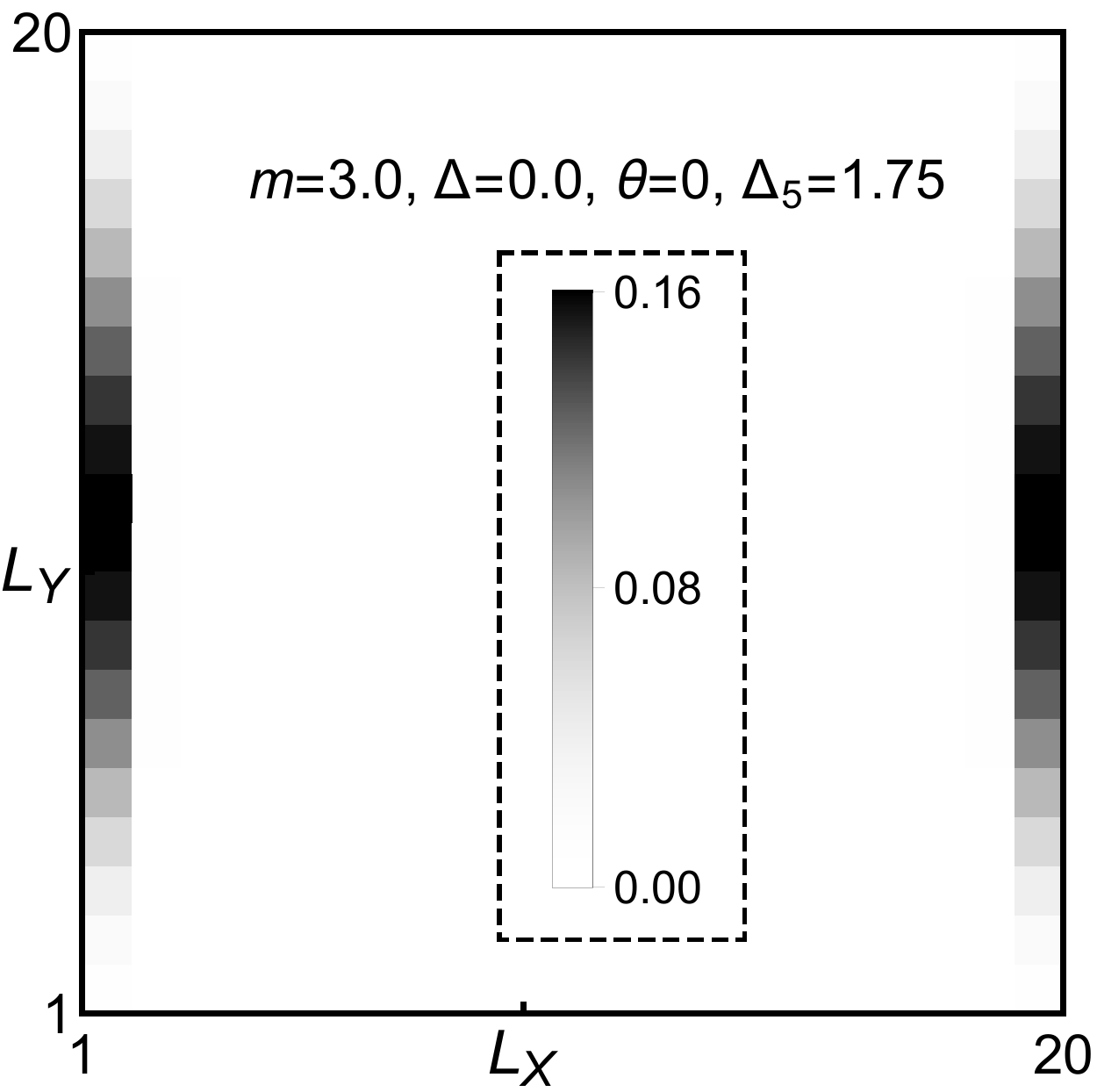}~\label{subfig1:delta5noDelta}}
\subfigure[]{\includegraphics[width=0.24\linewidth]{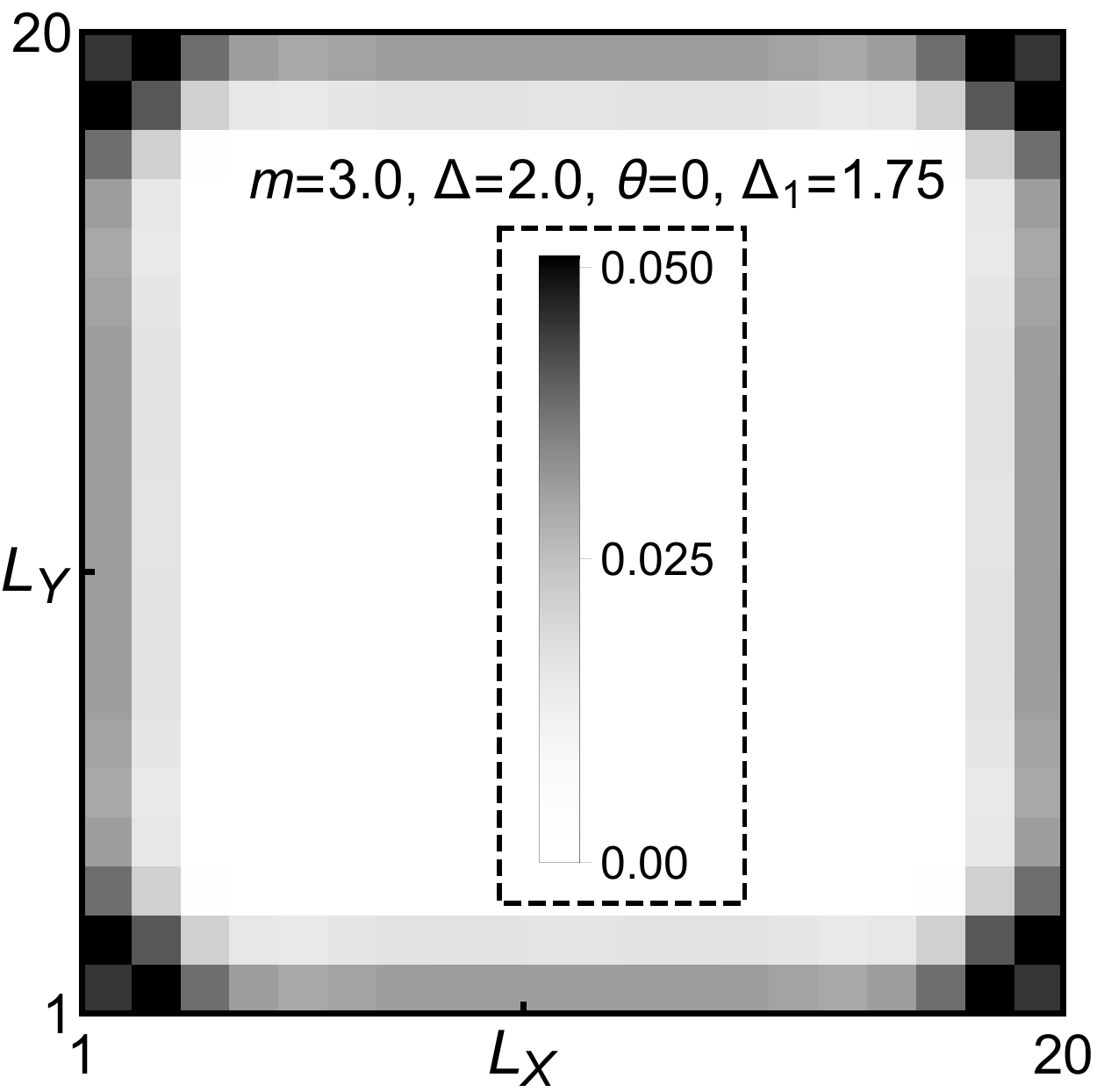}~\label{subfig1:delta1Delta}}%
\subfigure[]{\includegraphics[width=0.24\linewidth]{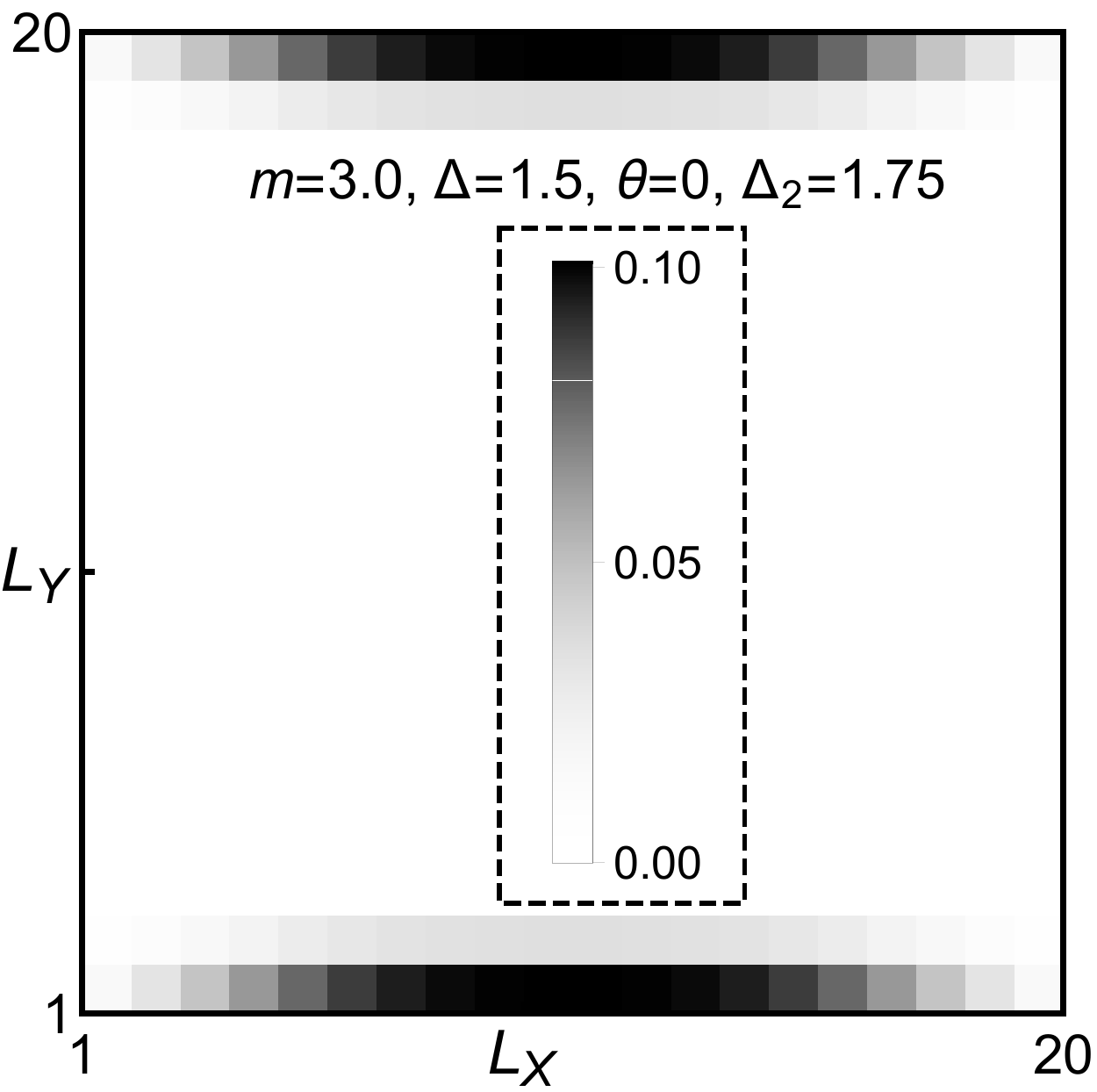}~\label{subfig1:delta2Delta}}%
\subfigure[]{\includegraphics[width=0.24\linewidth]{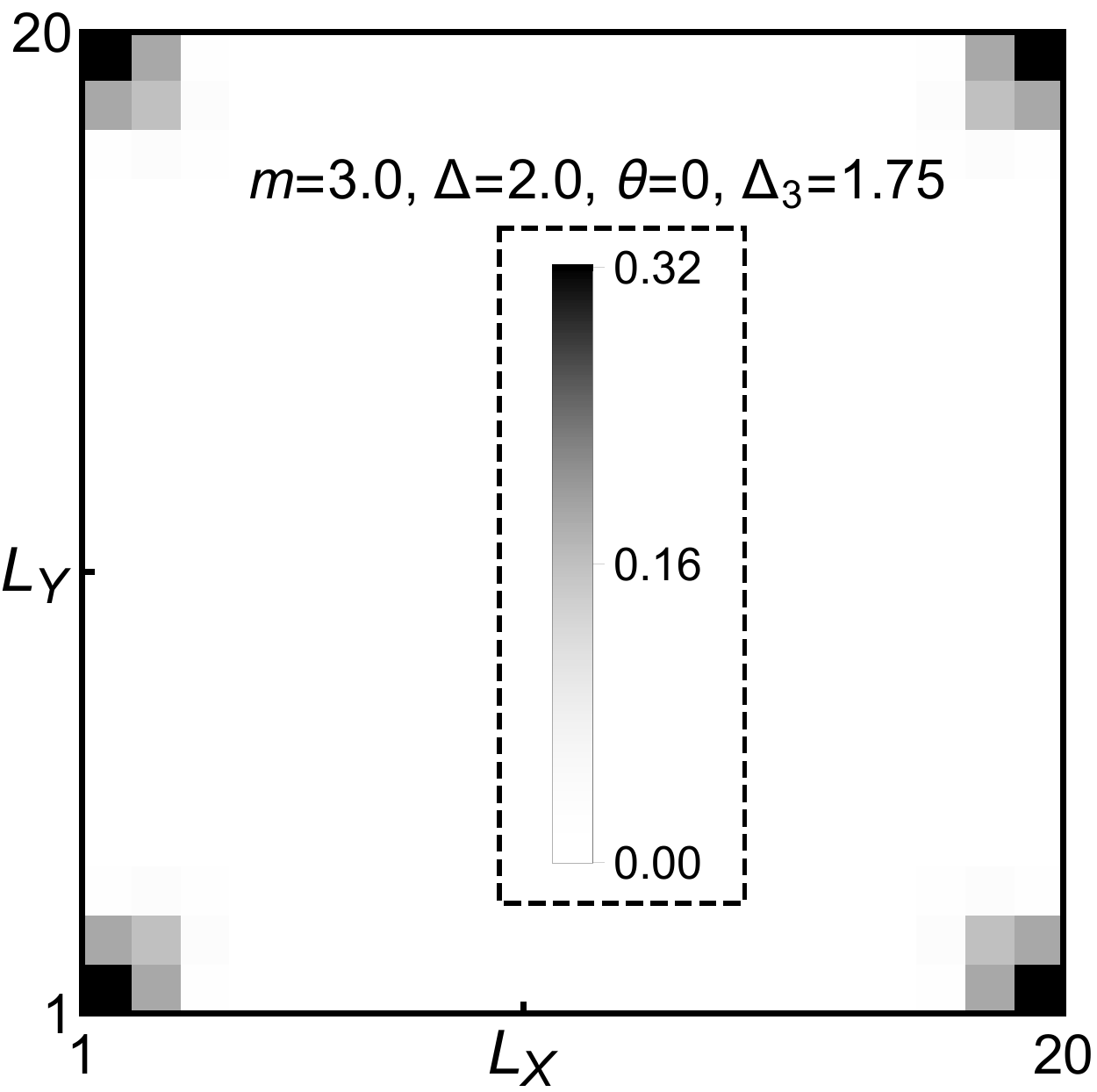}~\label{subfig1:delta3Delta}}%
\subfigure[]{\includegraphics[width=0.24\linewidth]{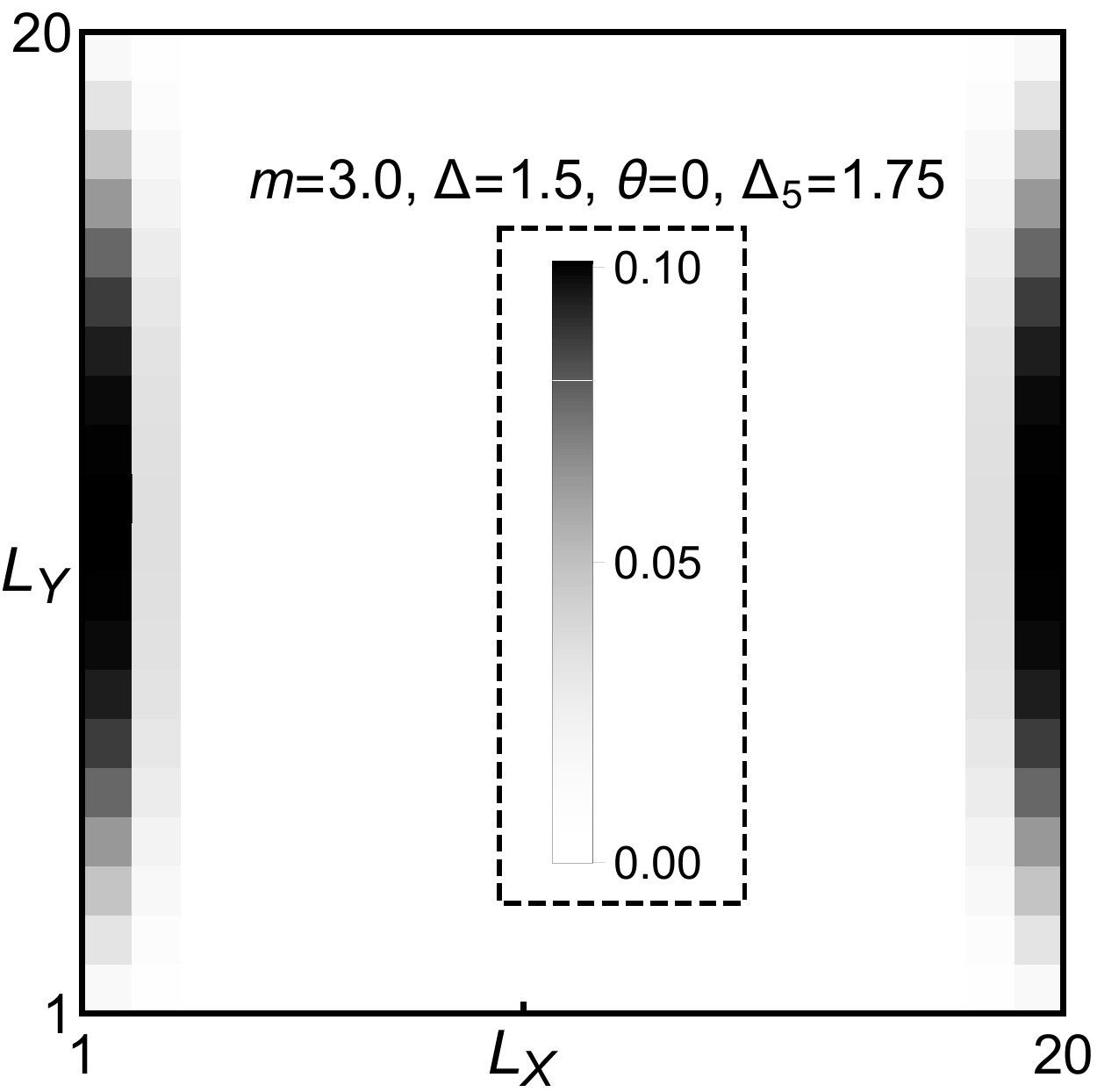}~\label{subfig1:delta5Delta}}
\caption{Local density of states (LDOS) for \emph{near} (due to finite system size) zero energy modes for various $p$-wave pairings, in the (a)-(d) absence or (e)-(h) presence of $C_4$ symmetry breaking WD mass $H_2(0)$. Here we set $t=t_0=1$, $m=3$ (yielding a NI) and $\mu=0$ (no Fermi surface). Only the $\Delta_3$ pairing anticommutes with $H_2(0)$ and represents a HOTSC, supporting four Majorana corner modes. Remaining three pairings commute with $H_2(0)$ and their boundary modes are mildy affected [Table~\ref{Table:algebra_2D}].    
}~\label{Fig:LDOS2D1}
\end{figure*}

\begin{figure}[t!]
\subfigure[]{\includegraphics[width=0.48\linewidth]{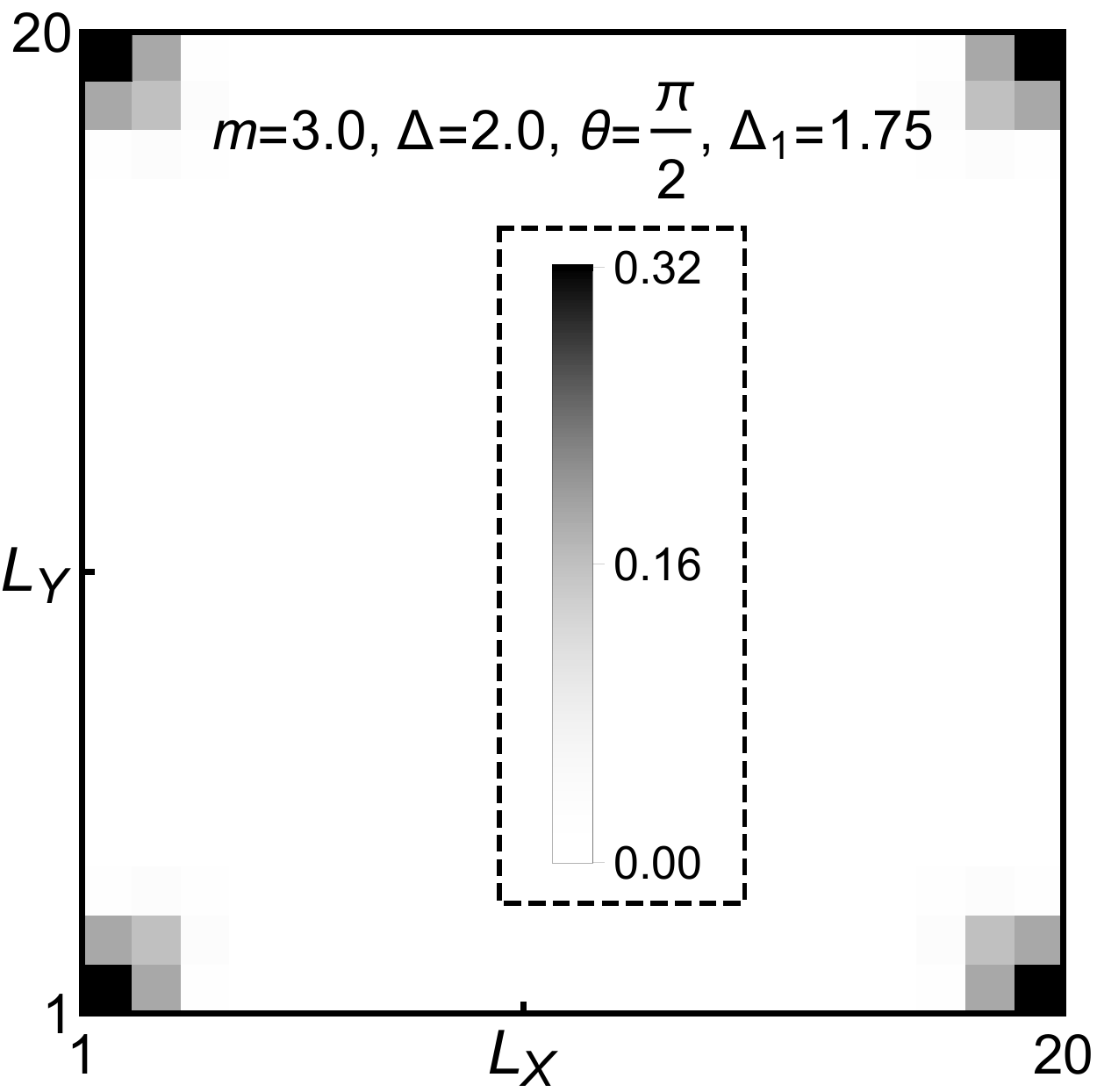}~\label{subfig:Delta_1Delta_2}}%
\subfigure[]{\includegraphics[width=0.48\linewidth]{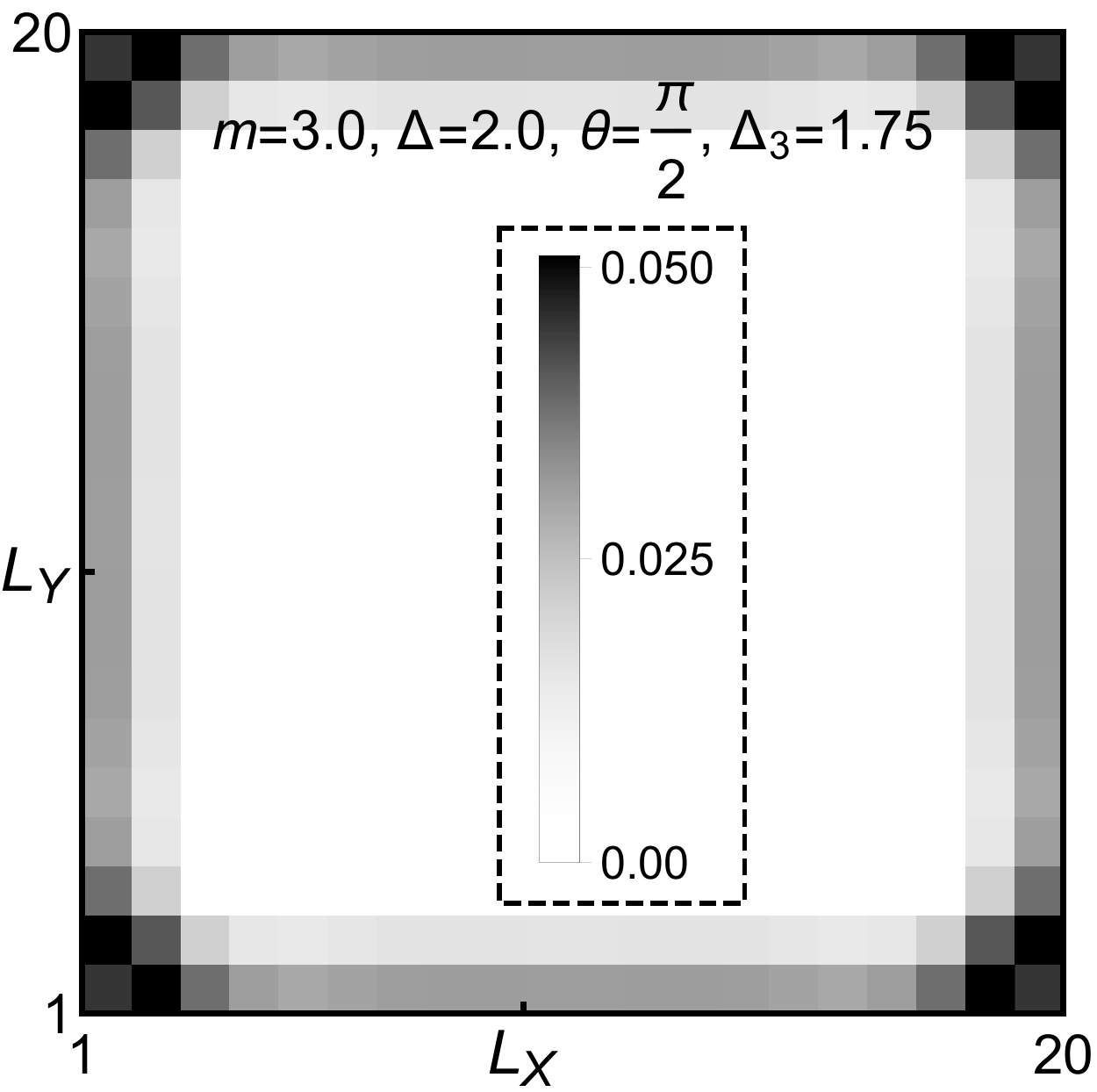}~\label{subfig:Delta_3Delta_2}}
\caption{LDOS for near zero energy modes for (a) $\Delta_1$ and (b) $\Delta_3$ pairing in the presence of $H_2(\frac{\pi}{2})$. Notice that $\Delta_1 (\Delta_3)$ pairing anticommutes (commutes) with $H_2(\frac{\pi}{2})$ [Table~\ref{Table:algebra_2D}], yielding a second- (first-)order TSC, supporting corner (edge) modes. 
}~\label{Fig:LDOS2D2}
\end{figure}

\begin{figure*}[t!]
\subfigure[]{\includegraphics[width=0.185\linewidth]{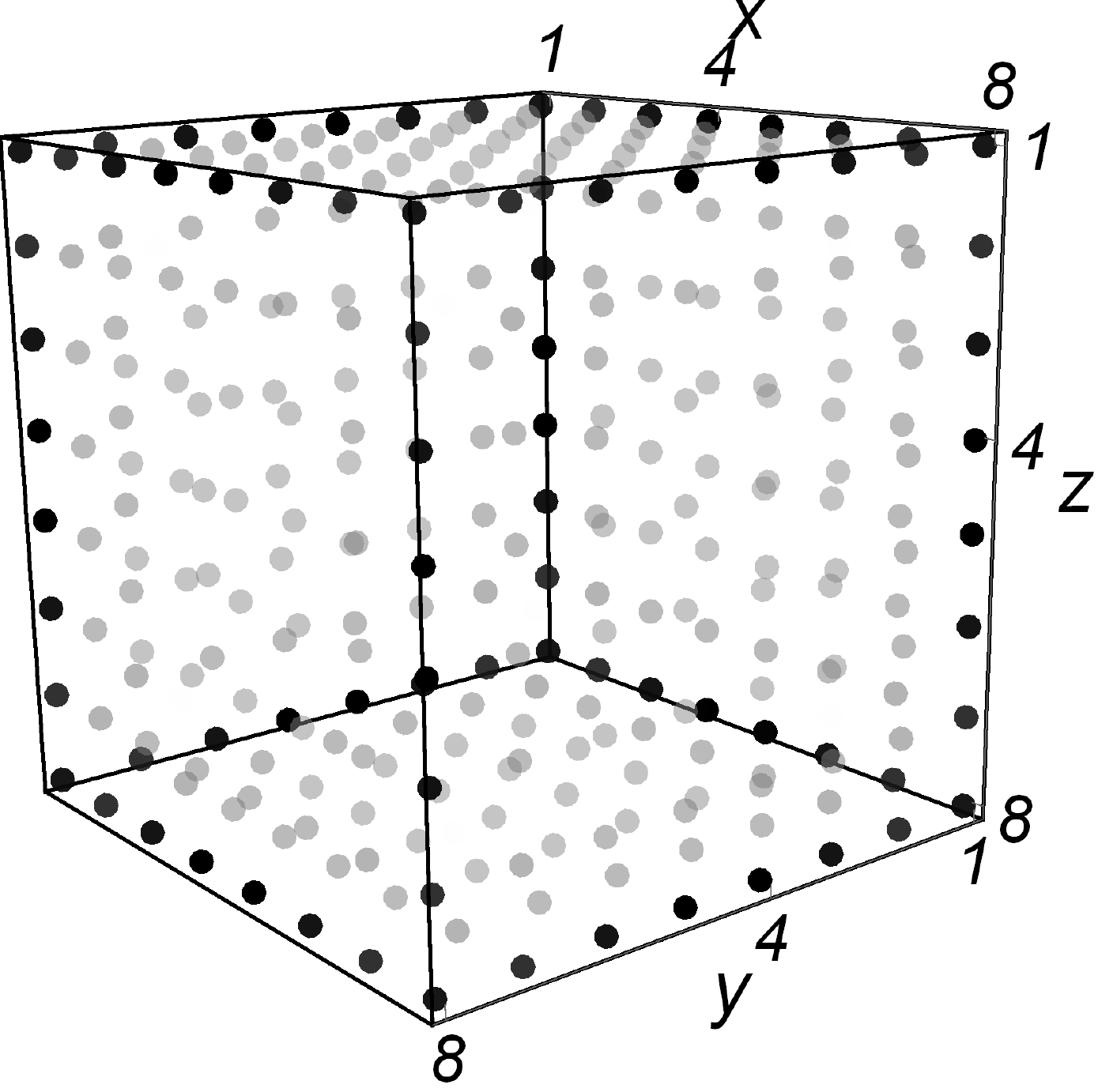}~\label{subfig:pwave_nodelta}}%
\includegraphics[width=0.055\linewidth]{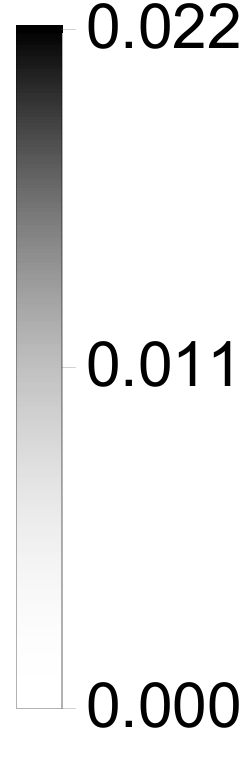}%
\subfigure[]{\includegraphics[width=0.185\linewidth]{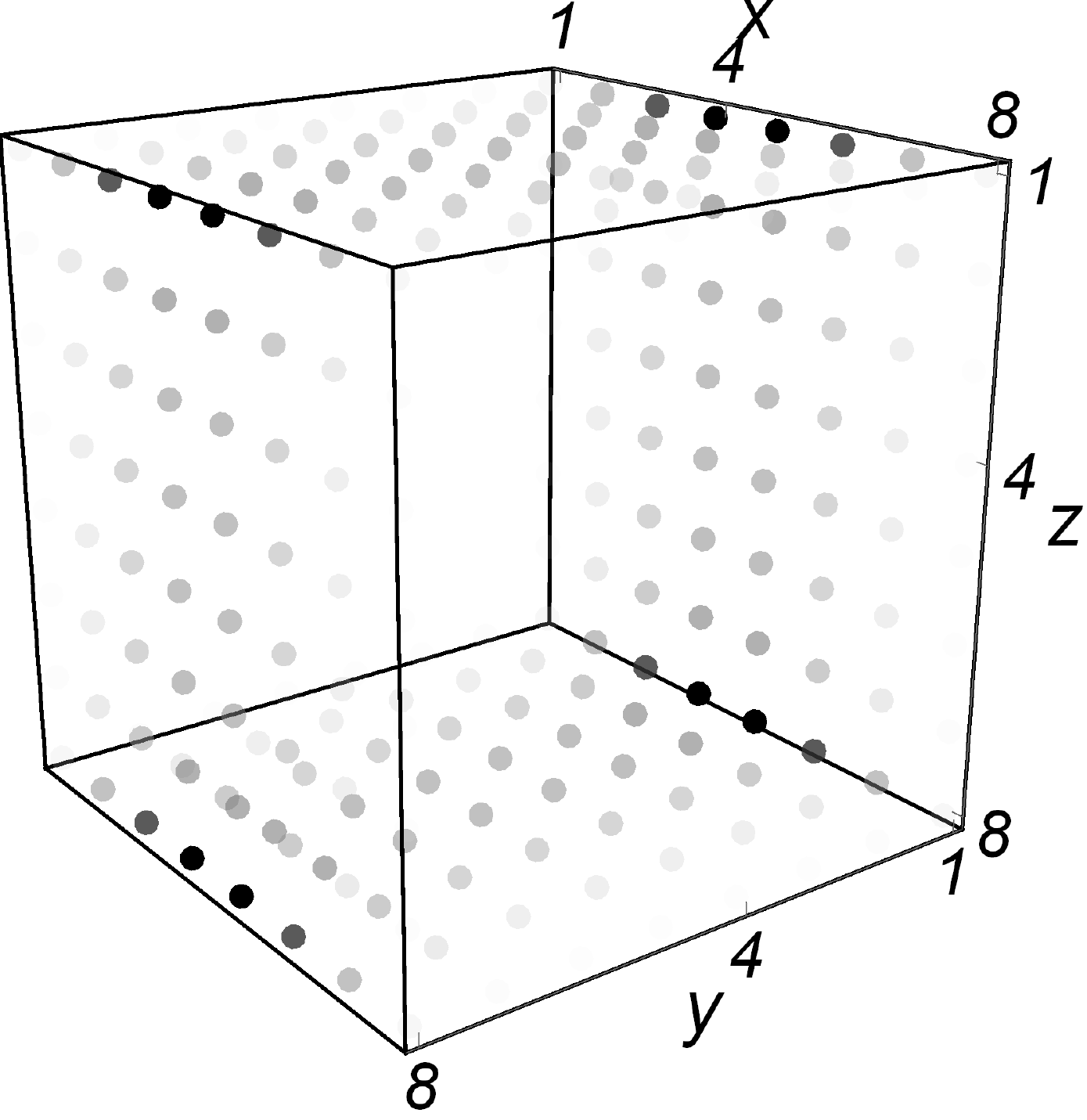}~\label{subfig:pxwave_nodelta}}%
\includegraphics[width=0.055\linewidth]{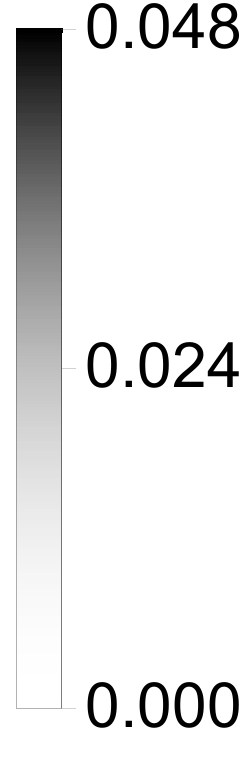}%
\subfigure[]{\includegraphics[width=0.185\linewidth]{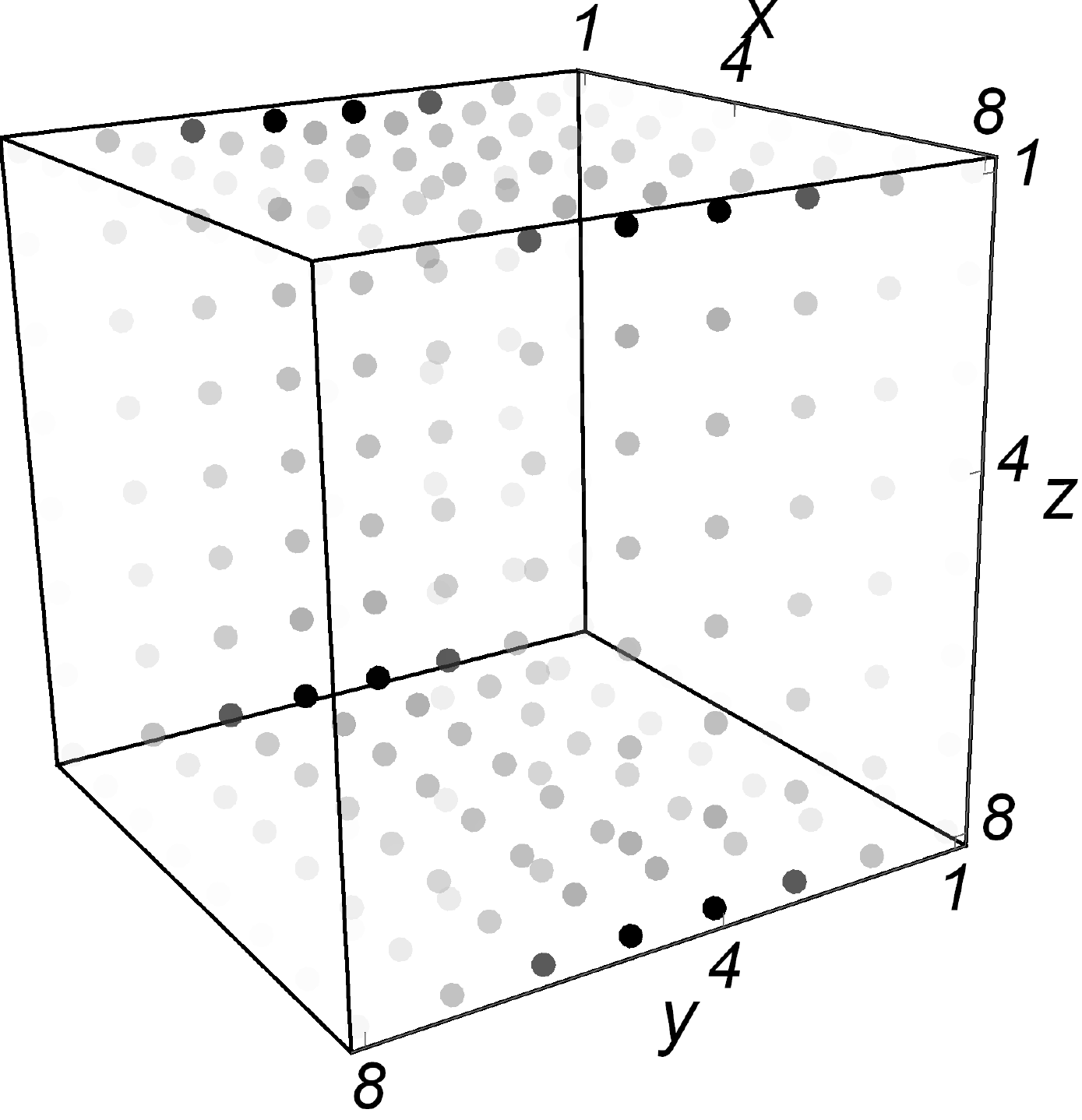}~\label{subfig:pywave_nodelta}}%
\includegraphics[width=0.055\linewidth]{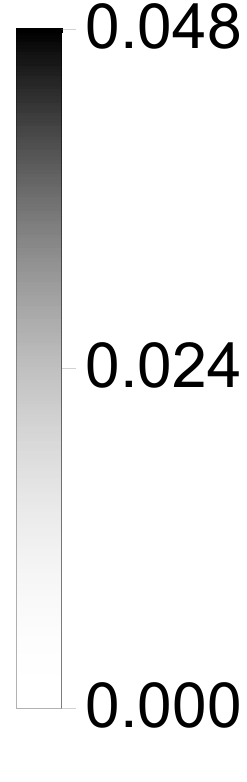}%
\subfigure[]{\includegraphics[width=0.185\linewidth]{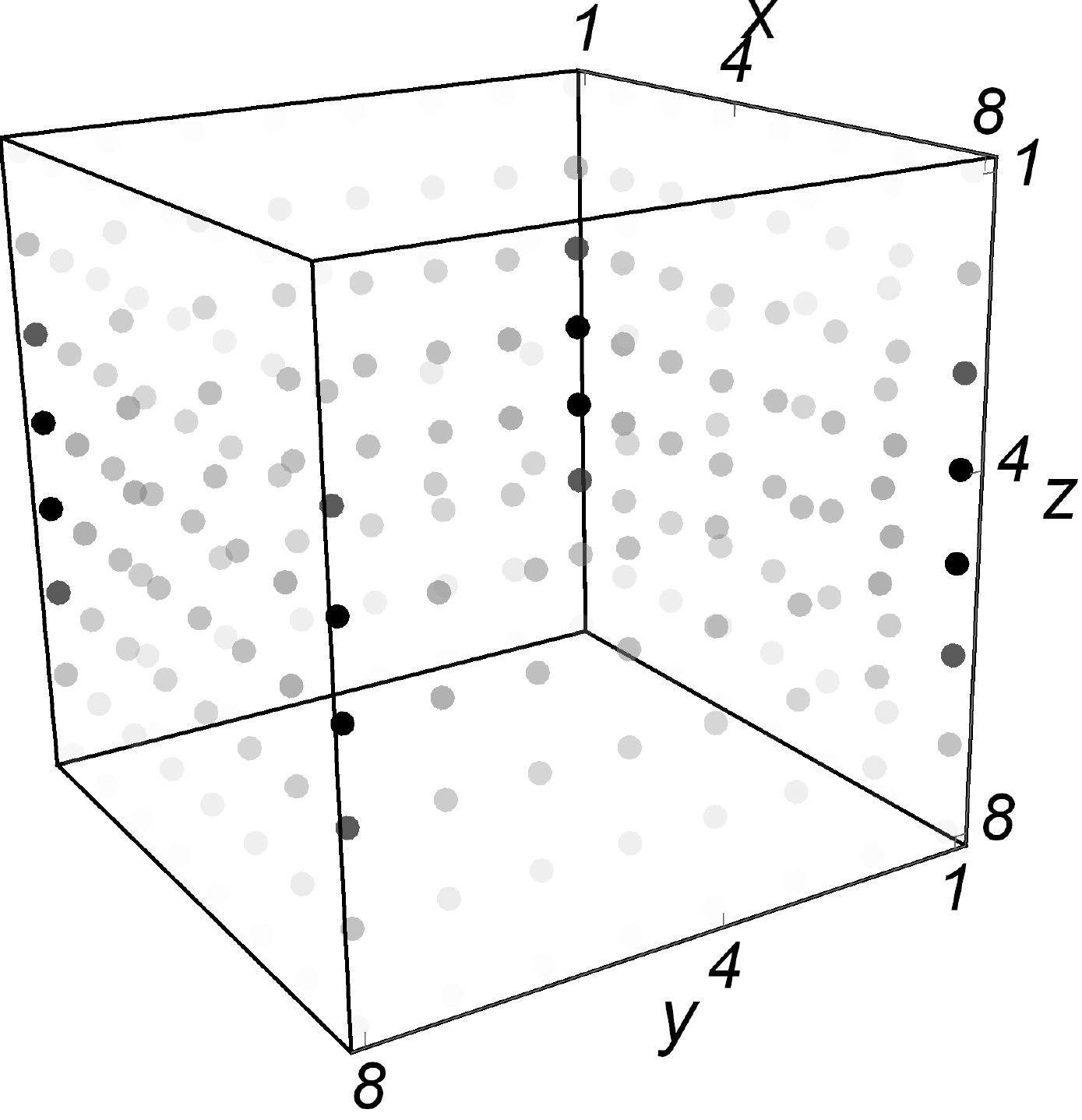}~\label{subfig:pzwave_nodelta}}%
\includegraphics[width=0.055\linewidth]{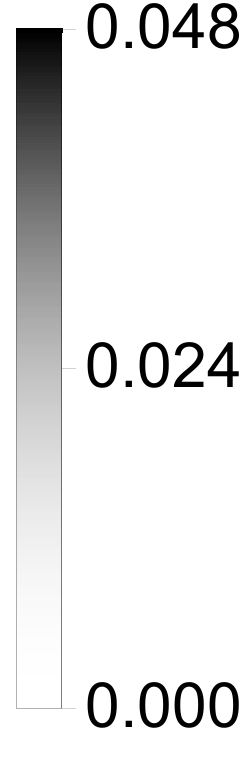}
\subfigure[]{\includegraphics[width=0.185\linewidth]{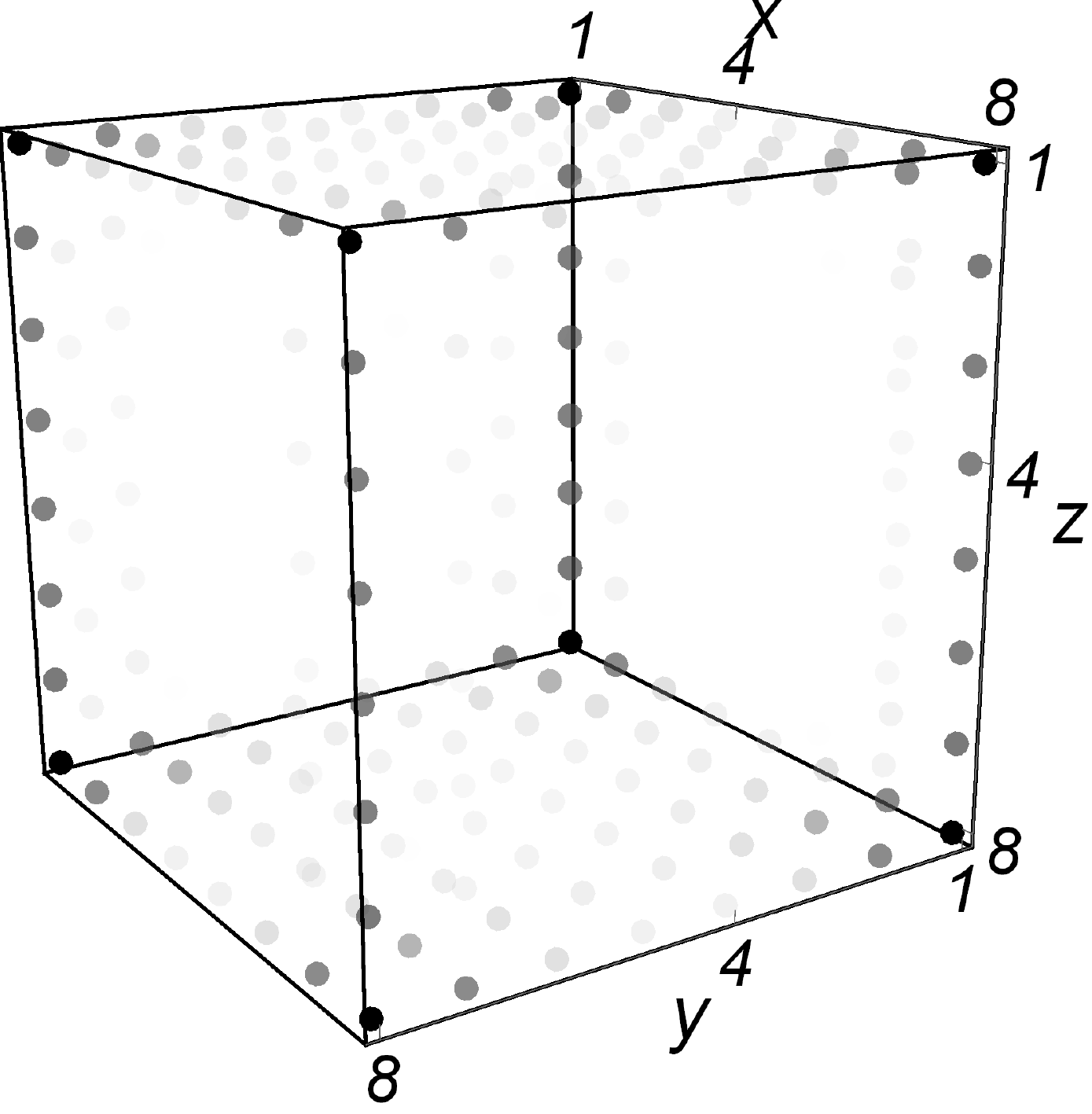}~\label{subfig:pwave_delta}}%
\includegraphics[width=0.055\linewidth]{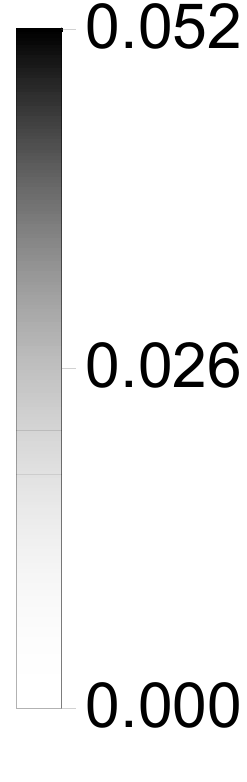}%
\subfigure[]{\includegraphics[width=0.185\linewidth]{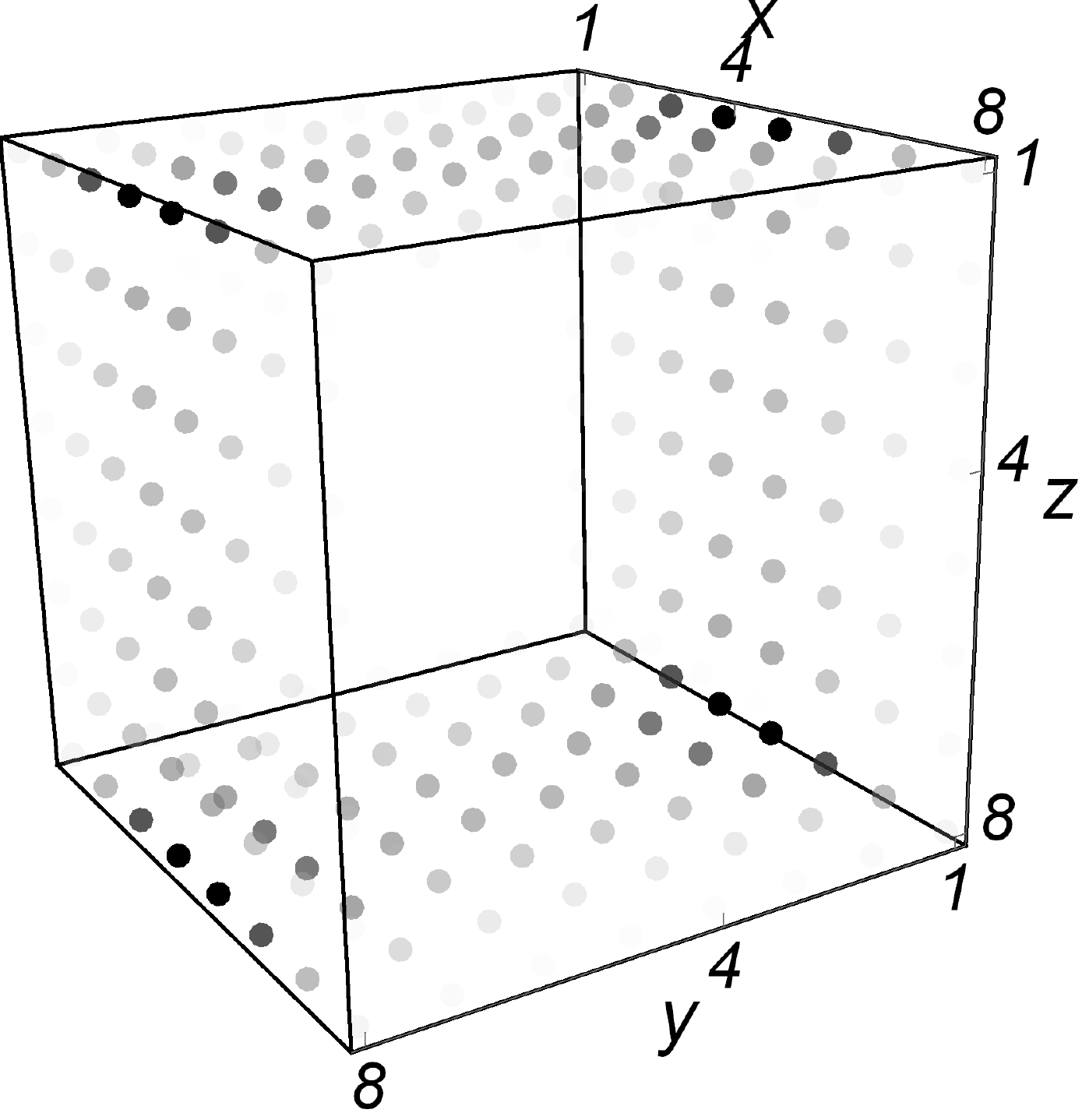}~\label{subfig:pxwave_delta}}%
\includegraphics[width=0.055\linewidth]{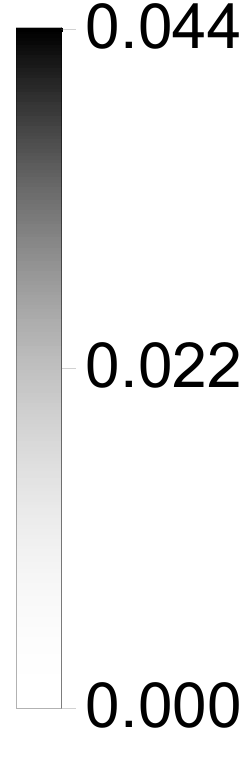}%
\subfigure[]{\includegraphics[width=0.185\linewidth]{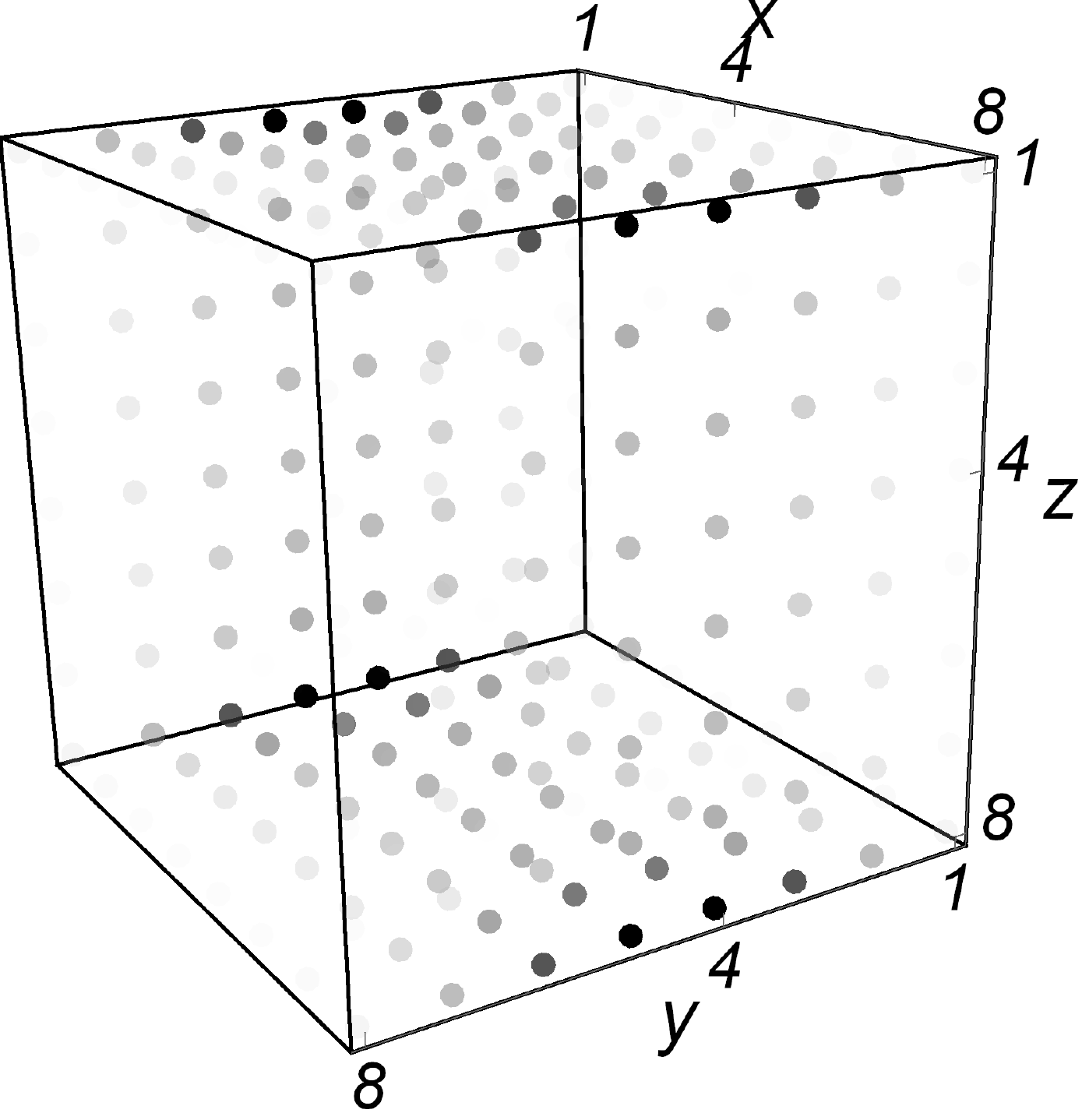}~\label{subfig:pywave_delta}}%
\includegraphics[width=0.055\linewidth]{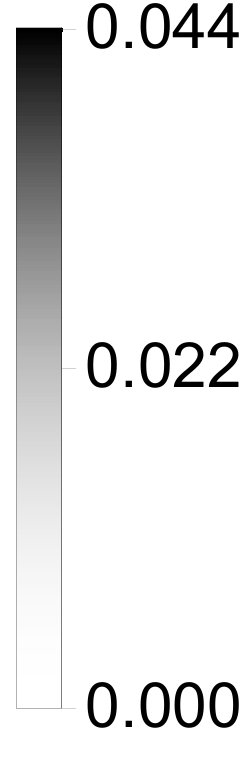}%
\subfigure[]{\includegraphics[width=0.185\linewidth]{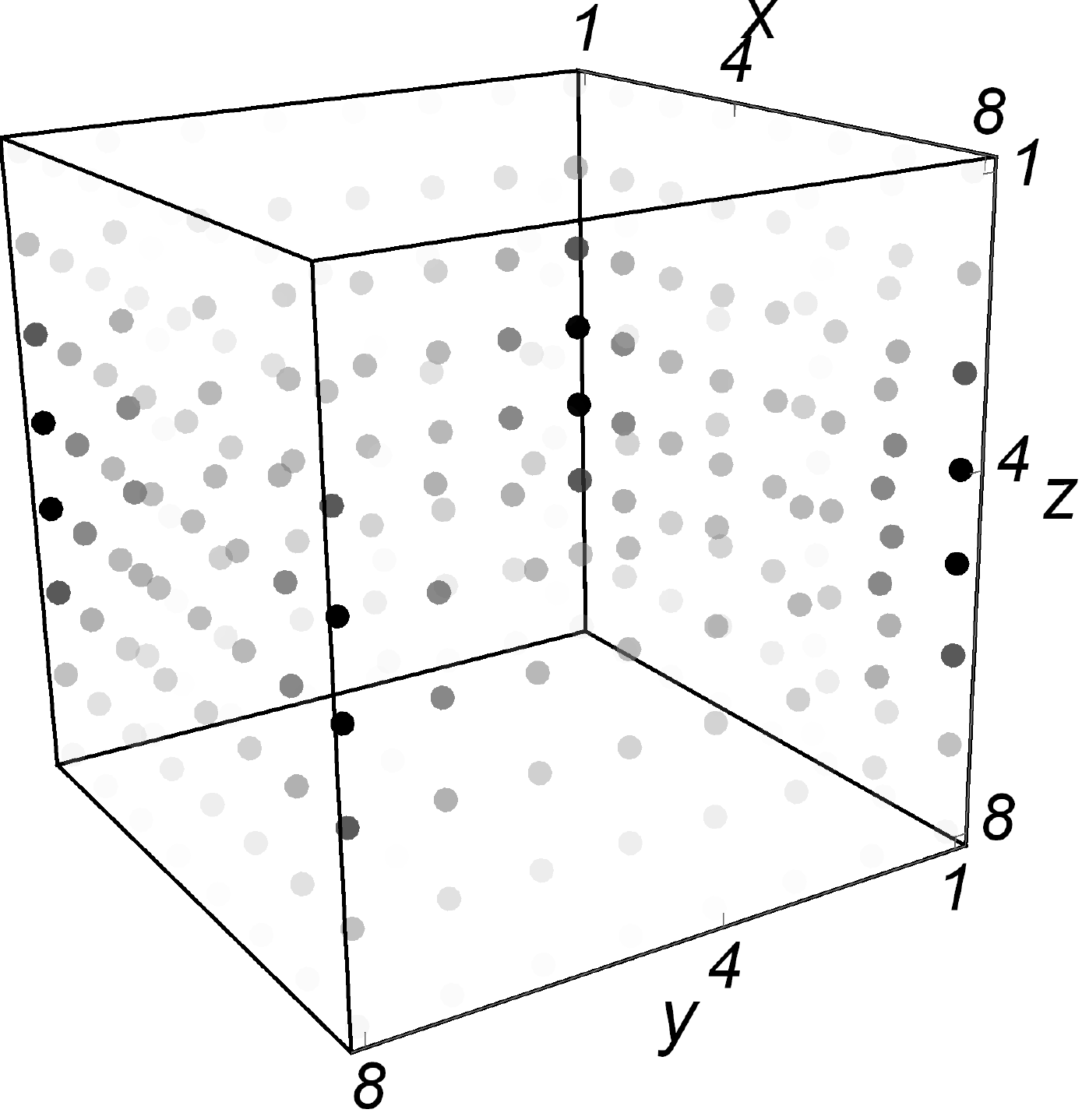}~\label{subfig:pzwave_delta}}%
\includegraphics[width=0.055\linewidth]{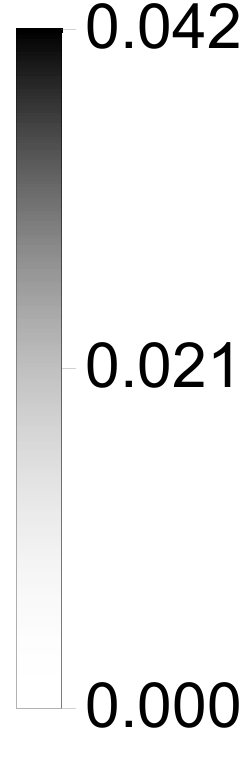}
\caption{LDOS for surface states of PS [(a),(e)], three components of vector pairing, $\Delta_1$ [(b),(f)], $\Delta_2$ [(c),(g)] and $\Delta_3$ [(d),(h)], in the (a)-(d) absence or (e)-(h) presence of the WD mass $H_2$. Only the PS pairing anticommutes with $H_2$ [Table~\ref{Table:algebra_3D}], and yields a HOTSC and hinge modes. Fermi arcs remain (almost) unchanged when $|\Delta|>0$, as all vector pairings commute with $H_2$. We set $\Delta=0$ (top) and $0.95$ (bottom), $t=t_0=1$, $m=4$ (producing trivial insulator), $\mu=0$, $\Delta_j=2$ for $j={\rm ps}, 1,2,3$.   
}~\label{Fig:LDOS3D}
\end{figure*}

In Dirac materials ${\mathcal P}$ and ${\mathcal T}$ can be simultaneously broken by a Dirac mass~\cite{peskin-schroeder, PTmass-latticecomment}. Here we show that both two- and three-dimensional Dirac systems allow one local or inter-unit cell or momentum independent pairing that \emph{anticommutes} with such a Dirac mass [Tables~\ref{Table:algebra_2D} and~\ref{Table:algebra_3D}]. In addition, when it breaks the $C_4$ symmetry (hereafter referred as the Wilson-Dirac (WD) mass) the boundary modes (with codimension $d_c=1$) of such paired states suffer dimensional reduction, producing zero energy corner and hinge modes ($d_c=2$) of neutral Majorana fermions, respectively in two and three dimensions [Figs.~\ref{Fig:LDOS2D1},~\ref{Fig:LDOS2D2} and~\ref{Fig:LDOS3D}]. The paired states then represent a higher-order topological superconductor (HOTSC), a topic of immense current interest~\cite{HOTreferencecoment,benalcazar2017,benalcazar-prb2017,song2017,schindler2018,matsugatani2018,schindler-sciadv2018,hsu2018,yan2018,trifunovic2019,xue2019,calugaru2019,fulga2019,ahn2018,agarwala2020,wang-lin-hughes-HOTSC,wu-yan-huang-HOTSC,wang-liu-lu-zhang-HOTSC,liu-he-nori-HOTSC,Klinovaja-HOTSC-1,yan-HOTSC,cole-HOTSC,zhu-HOTSC,pan-yang-chen-xu-liu-liu-HOTSC,ghorashi-HOTSC,fulga-HOTSC-1,trauzettel-HOTSC,bjyang-HOTSC-1,dassarma-HOTSC,srao-HOTSC,bomantara-HOTSC}. By contrast, other topological $p$-wave pairings, which \emph{commute} with the WD mass, continue to represent first-order TSCs.

We primarily support these outcomes by \emph{numerically} diagonalizing the effective single-particle Hamiltonian for various local pairings in \emph{trivial} Dirac insulators (devoid of any boundary modes) and in the absence of a Fermi surface (thus involving both intra- and inter-band couplings). Therefore, emergent topology can solely be attributed to the Bogoliubov-de Gennes (BdG) quasiparticles~\cite{comment-topology-NS}. In the presence of a Fermi surface (conforming to weak coupling pairing) the \emph{intraband} component of a HOTSC mimics $p+id$ pairing. Respectively, the parity mixing and ${\mathcal T}$-breaking arise from the lack of ${\mathcal P}$ and ${\mathcal T}$ symmetries, whereas the $d$-wave component roots into the $C_4$ symmetry breaking in the NS, as its restoration converts the paired state into a trivial $p+is$ pairing. Therefore, antiferromagnetic topological insulators~\cite{PTmass-latticecomment,moore-AFTI}, such as  MnBi$_2$Te$_4$~\cite{magnetic-TI-1}, can be the ideal platform to search for HOTSCs, when they are \emph{doped} (yielding a pairing conducive Fermi surface) and \emph{strained} (lifting the crystalline symmetry).

\emph{Two-dimensions}.~The Hamiltonian for 2D quadrupolar massive Dirac fermions is $H^{\rm 2D}_{\rm Dir}=H_0 + H_1 + H_2 (\theta)$, where
\allowdisplaybreaks[4]
\begin{eqnarray}~\label{Eq:quadrupolarDirac_2D}
H_0 &=& t \left[ \Gamma_{331} S_1 + \Gamma_{302} S_2 \right], 
H_1= \Gamma_{303} \big[ m + t_0 \sum^2_{j=1} C_j \big], \nonumber \\
H_2 (\theta) &=& \Delta \; g({\bf k}) \; \left[ \Gamma_{011} \cos\theta + \Gamma_{021} \sin\theta \right].
\end{eqnarray}
Here $S_j=\sin(k_j a)$, $C_j=\cos(k_j a)$ with $k_j$s as the components of spatial momenta, $a$ is the lattice spacing, and $\theta$ is an arbitrary parameter~\cite{theta-comment}. Eight-dimensional Hermitian matrices $\Gamma_{\lambda \nu \rho}=\eta_\lambda \sigma_\nu \tau_\rho$, where $\lambda, \nu, \rho=0,\cdots, 3$, and the Pauli matrices $\{ \tau_\nu \}$, $\{ \sigma_\nu \}$ and $\{ \eta_\nu \}$ operator on the parity, spin and particle-hole indices, respectively. The Nambu spinor basis is $\Psi^\top_{\rm Nam} ({\bf k})= \big[ \Psi_{\bf k}, \sigma_2 \tau_0 ( \Psi^\dagger_{-\bf k} )^\top \big]$, where  $\Psi^\top_{\bf k}= \big[ c^+_{{\bf k},\uparrow}, c^-_{{\bf k}, \uparrow}, c^+_{{\bf k},\downarrow}, c^-_{{\bf k},\downarrow} \big]$, and $c^\tau_{{\bf k},\sigma}$ is fermion annihilation operator with parity $\tau=\pm$, spin projection $\sigma=\uparrow,\downarrow$ and momentum ${\bf k}$. When $\Delta=0$, $H^{\rm 2D}_{\rm Dir}$ describes a quantum spin Hall insulator and a trivial or normal insulator (NI) for $|m/t_0|<2$ and $|m/t_0|>2$, respectively. Here we set $m/t_0=3$.

Notice that $H_2(\theta)$ is \emph{odd} under ${\mathcal T}: {\bf k} \to -{\bf k}$ and ${\mathcal T}=\Gamma_{020}{\mathcal K}$, where ${\mathcal K}$ is the complex conjugation, and ${\mathcal P}: {\bf r} \to -{\bf r}$ and ${\mathcal P} \Psi_{\rm Nam}({\bf k}) {\mathcal P}^{-1} = \Gamma_{003} \Psi_{\rm Nam}(-{\bf k})$. In addition,  $H_2(\theta)$ breaks $C_4$ symmetry, if $g ({\bf k})=C_1-C_2$. Then it represents a WD mass, as $\{ H_2(\theta), H_1+H_2 \}=0$, and the excitations are massive \emph{quadrupolar} Dirac fermions.

Now we address topology of the paired states in this system, described by the single-particle Hamiltonian 
\allowdisplaybreaks[4]
\begin{eqnarray}~\label{Eq:localpairing_2D}
H^{\rm 2D}_{\rm pair} &=& \Delta_1 \Gamma_{j22} + \Delta_2 \Gamma_{j32} + \Delta_3 \Gamma_{j12} +\Delta_4 \Gamma_{j00} \nonumber \\
&+& \Delta_5 \Gamma_{j01} + \Delta_6 \Gamma_{j03},
\end{eqnarray}
where $j=1$ or $2$, reflecting the $U(1)$ gauge freedom in the choice of superconducting phase, and $\Delta_l$'s are the pairing amplitudes. We choose $j=1$. Two $s$-wave pairings ($\Delta_4,\Delta_6$) are topologically trivial. By contrast, $\Delta_1$ and $\Delta_3$ pairings fully anticommute with the Dirac kinetic energy $H_0$ and represent fully gapped topological superconductor when $|\Delta_j|> |m-2 t_0|$ and $\Delta=0$. They support one-dimensional (1D) edge modes [Figs.~\ref{subfig1:delta1noDelta} and~\ref{subfig1:delta3noDelta}]. The remaining two pairings ($\Delta_2$ and $\Delta_5$) break rotational symmetry. Respectively, they produce two Dirac points along the $k_x$ and $k_y$ directions, and support Fermi arcs along the $x$ [Fig.~\ref{subfig1:delta2noDelta}] and $y$ [Fig.~\ref{subfig1:delta5noDelta}] directions.

\begin{table}[]
\begin{tabular}{|c|c|c|c|c|c|c| c|c|}
\hline
\multirow{2}{*}{Pairing} & \multirow{2}{*}{Matrix} & \multicolumn{2}{c|}{$H_0$} & \multirow{2}{*}{$H_1$} & \multirow{2}{*}{$H_2(0)$} & \multirow{2}{*}{$H_2\left( \frac{\pi}{2}\right)$} & \multicolumn{2}{c|}{HOTSC} \\ \cline{3-4} \cline{8-9}
   &  & $\Gamma_{331}$ & $\Gamma_{302}$ &  &  &  & $\theta=0$ & $\theta=\frac{\pi}{2}$ \\ \hline \hline
 $\Delta_1$ & $\Gamma_{j22}$ & $-$ & $-$ & $+$ & $+$ & $-$ & \xmark & \cmark \\ \hline 
 $\Delta_2$ & $\Gamma_{j32}$ & $+$ & $-$ & $+$ & $+$ & $+$ & \xmark & \xmark \\ \hline
 $\Delta_3$ & $\Gamma_{j12}$ & $-$ & $-$ & $+$ & $-$ & $+$ & \cmark & \xmark \\ \hline
 $\Delta_4$ & $\Gamma_{j00}$ & $-$ & $-$ & $-$ & $+$ & $+$ & \xmark & \xmark \\ \hline
 $\Delta_5$ & $\Gamma_{j01}$ & $-$ & $+$ & $+$ & $+$ & $+$ & \xmark & \xmark \\ \hline
 $\Delta_6$ & $\Gamma_{j03}$ & $+$ & $+$ & $-$ & $-$ & $-$ & \xmark & \xmark \\ \hline
\end{tabular}
\caption{Commutation ($+$)/anticommutation ($-$) relations of pairings [Eq.~(\ref{Eq:localpairing_2D})] with various entries of $H^{\rm 2D}_{\rm Dir}$ [Eq.~(\ref{Eq:quadrupolarDirac_2D})]. Two $s$-wave pairings ($\Delta_4,\Delta_6$) are topologically trivial. Remaining ones correspond to $p$-wave pairings. The associated boundary modes are shown in Fig.~\ref{Fig:LDOS2D1} (Top). When $\theta=0 (\frac{\pi}{2})$, only $\Delta_3 (\Delta_1)$ pairing \emph{anticommutes} with the WD mass $H_2(\theta)$ and supports four corner modes [Figs.~\ref{Fig:LDOS2D1} (bottom) and~\ref{Fig:LDOS2D2}]. 
}~\label{Table:algebra_2D}
\end{table}

The effect of the WD mass on the pairings and boundary modes can be anticipated from its (anti)commutation relations with the pairing matrices [Table~\ref{Table:algebra_2D}]. Here we consider $\theta=0$ and $\theta=\frac{\pi}{2}$ only. The $\Delta_3 [\Delta_1]$ pairing \emph{anticommutes} with $H_2(0) [H_2(\frac{\pi}{2})]$. As $H_2(\theta)$ changes sign under four-fold rotation, it acts as a \emph{domain wall mass} (see Ref.~\cite{Rackiw-Rebbi}) for the one-dimensional counter propagating edge modes of $\Delta_1 (\Delta_3)$ pairing when $\theta=0(\frac{\pi}{2})$. Consequently, the boundary modes of $\Delta_3$ pairing undergoes a dimensional reduction, yielding four corner localized Majorana zero modes with $d_c=2$ when $\theta=0$ [Fig.~\ref{subfig1:delta3Delta}]. And we realize a second-order TSC. By contrast, $\Delta_1$, $\Delta_2$ and $\Delta_5$ pairings commute with $H_2(0)$. So, their boundary modes are \emph{mildly} affected [Figs.~\ref{subfig1:delta1Delta},~\ref{subfig1:delta2Delta} and~\ref{subfig1:delta5Delta}], and they continue to be first-order TSCs. Similarly, for $\theta=\frac{\pi}{2}$, only the $\Delta_1$ pairing represents a HOTSC and supports four corner localized zero modes [Fig.~\ref{Fig:LDOS2D2}].

Based on this observation, we propose a \emph{general principle} of realizing HOTSCs. If an effective single-particle Hamiltonian ($H_{\rm FO}$) describes a first-order TSC, supporting boundary modes of $d_c=1$, then the addition of an appropriate discrete-symmetry breaking WD mass ($H_{\rm WD}$) can covert the paired state into a HOTSC, when $\left\{ H_{\rm FO}, H_{\rm WD} \right\}=0$. Next we show that this mechanism is operative in $d=3$, yielding 1D hinge modes of $d_c=2$.

The Hamiltonian for a collection of three-dimensional quadrupolar massive Dirac fermions is $H^{\rm 3D}_{\rm Dir}=H_0+ H_1+H_2$, where $H_2 = \Gamma_{020} \Delta \; g({\bf k})$ is the WD mass, 
\allowdisplaybreaks[4]
\begin{eqnarray}~\label{Eq:quadrupolarDirac_3D}
H_0 = t \sum^{3}_{j=1} \Gamma_{31j} S_j, \:
H_1= \Gamma_{330} \big[m- t_0 \sum^{3}_{j=1} C_j \big],
\end{eqnarray}
and $\Gamma_{\rho \lambda \nu}=\eta_\rho \tau_\lambda \sigma_\nu$. The spinor basis is $\Psi^\top_{\rm Nam} ({\bf k})=\big[ \Psi_{\bf k}, \tau_0 \sigma_2 ( \Psi^\dagger_{-{\bf k}} )^\top \big]$, with $\Psi^\top_{\bf k} =[c^+_{{\bf k},\uparrow}, c^+_{{\bf k},\downarrow}, c^-_{{\bf k},\uparrow}, c^-_{{\bf k},\downarrow}]$. While $H_0$ gives rise to massless Dirac fermions, $H_1$ stands as the symmetry preserving Dirac mass. Under ${\mathcal T}:{\bf k} \to -{\bf k}$ and $\Psi_{\rm Nam}({\bf k}) \to \Gamma_{002} \Psi_{\rm Nam}({-{\bf k}})$, and the corresponding antiunitary operator is ${\mathcal T}=\Gamma_{002} {\mathcal K}$. Under ${\mathcal P}: {\bf r} \to -{\bf r}$ and ${\mathcal P} \Psi_{\rm Nam}({\bf k}) {\mathcal P}^{-1} =\Gamma_{330} \Psi_{\rm Nam}(-{\bf k})$. The WD mass $H_2$ breaks both ${\mathcal P}$ and ${\mathcal T}$ symmetries, but preserves the conjugate ${\mathcal P}{\mathcal T}$ symmetry, and in addition also breaks the discrete $C_4$ symmetry if $g({\bf k})=C_1-C_2$.

For $\Delta=0$, the above model supports a strong (weak) topological insulator for $1<m/t_0<3$ ($-1<m/t_0<1$), and NIs for $m/t_0>3$ and $m/t_0<-1$~\cite{ohtsuki-TImodel}. Here we choose $m/t_0=4$. So the Dirac insulator is always trivial, and we solely unveil the topology of the paired states.

The single-particle Hamiltonian for local pairings is 
\allowdisplaybreaks[4]
\begin{eqnarray}~\label{Eq:singleparticleH_Pairing_3D}
H^{\rm 3D}_{\rm pair}= \Delta_{\rm s} \Gamma_{j00} + \Delta_{\rm ps} \Gamma_{j10} + \Delta_0 \Gamma_{j30} + \sum^3_{i=1} \Delta_i \Gamma_{j2i},
\end{eqnarray}  
where $j=1$ or $2$. In the Dirac language $\Delta_{\rm s}$ transforms as a scalar, while the odd-parity $\Delta_{\rm ps}$ transforms as a \emph{pseudoscalar} (PS). On the other hand, $\Delta_0$ and $\Delta_i$ transform as the temporal and three spatial components ($i=1,2,3$) of vector pairing, respectively~\cite{ohsaku}. The (anti)commutation relations of all pairings with various components of $H^{\rm 3D}_{\rm Dir}$ are shown in Table~\ref{Table:algebra_3D}, which determines their topology. First we set $\Delta=0$.

\begin{table}[t!]
\begin{tabular}{|c|c|c|c|c|c|c|c|}
\hline
\multirow{2}{*}{Pairing} & \multirow{2}{*}{Matrix} & \multicolumn{3}{c|}{$H_0$} & \multirow{2}{*}{$H_1$} & \multirow{2}{*}{$H_2$} & \multirow{2}{*}{HOTSC}  \\ \cline{3-5} 
 &  & $\Gamma_{311}$ & $\Gamma_{312}$ & $\Gamma_{313}$ &  &  &  \\ \hline \hline
$\Delta_s$        & $\Gamma_{j00}$ & $-$ & $-$ & $-$ & $-$ & $+$ & \xmark \\ \hline
$\Delta_{\rm ps}$ & $\Gamma_{j10}$ & $-$ & $-$ & $-$ & $+$ & $-$ & \cmark \\ \hline
$\Delta_0$        & $\Gamma_{j30}$ & $+$ & $+$ & $+$ & $-$ & $-$ & \xmark \\ \hline
$\Delta_1$        & $\Gamma_{j21}$ & $+$ & $-$ & $-$ & $+$ & $+$ & \xmark \\ \hline
$\Delta_2$        & $\Gamma_{j22}$ & $-$ & $+$ & $-$ & $+$ & $+$ & \xmark \\ \hline
$\Delta_3$        & $\Gamma_{j23}$ & $-$ & $-$ & $+$ & $+$ & $+$ & \xmark \\ \hline
\end{tabular}
\caption{Commutation ($+$)/anticommutation ($-$) relations of local pairings [Eq.~(\ref{Eq:singleparticleH_Pairing_3D})] with various components of $H^{\rm 3D}_{\rm Dir}$ [Eq.~(\ref{Eq:quadrupolarDirac_3D})]. Two $s$-wave pairings ($\Delta_s$ and $\Delta_0$) are topologically trivial. Only the pseudoscalar (PS) pairing ($\Delta_{\rm ps}$) anticommutes with the WD mass ($H_2$), yielding a second-order TSC that supports one-dimensional hinge modes, see Fig.~\ref{subfig:pwave_delta}. 
}~\label{Table:algebra_3D}
\end{table}

Two $s$-wave pairings $\Delta_{\rm s}$ and $\Delta_{\rm 0}$ are topologically trivial. The PS pairing is a fully gapped class DIII TSC, supporting gapless Majorana modes on six surfaces of a cubic system [Fig.~\ref{subfig:pwave_nodelta}], when $|\Delta_{\rm ps}|>|m-3t_0|$~\cite{fu-berg}. The spatial components of vector pairing break the rotational symmetry and represent nematic superconductors~\cite{liangfu-nematicSC}. The $\Delta_i$ pairing supports two Dirac points along the $k_i$ direction~\cite{goswami-roy-axionFT}, and Fermi arc surface states (with $d_c=1$) along the $i$th direction in the real space [Figs.~\ref{subfig:pxwave_nodelta}-~\ref{subfig:pzwave_nodelta}].

Now we switch on the WD mass $H_2$. It breaks the $C_4$ symmetry and changes sign four times in the $xy$ plane for any $z$. The PS pairing anticommutes with $H_2$ [Table~\ref{Table:algebra_3D}], which then acts as a domain wall mass for the surface Majorana fermions, yielding 1D hinge modes (with $d_c=2$) along the $z$ direction [Fig.~\ref{subfig:pwave_delta}]. The PS pairing then represents a second-order TSC. For $g({\bf k})=C_3-C_1$ and $C_2-C_3$ the hinge modes are aligned along the $y$ and $x$ directions, respectively. By contrast, all vector pairings commute with $H_2$, and thus the Fermi arcs are mildly affected [Figs.~\ref{subfig:pxwave_delta}-~\ref{subfig:pzwave_delta}]. These outcomes are in agreement with the \emph{general principle} of realizing HOTSC.

Finally, we anchor these findings by projecting the HOTSCs onto a Fermi surface when the chemical doping (here measured from the bottom of the conduction band) $\mu>0$. We return to two-dimensional Dirac system and consider its low-energy model for $\theta=0$
\allowdisplaybreaks[4]
\begin{equation}
H^{\rm 2D, low}_{\rm Dir}=v \left[ \Gamma_{331} k_1 + \Gamma_{302} k _2 \right] + \Gamma_{303} M + \Gamma_{011} \bar{\Delta} (k^2_1-k^2_2), 
\end{equation}
obtained by expanding $H^{\rm 2D}_{\rm Dir}$ around the $\Gamma=(0,0)$ point, where $M=|m - 2t_0|$, $v = t a$ bears the dimension of the Fermi velocity, and $\bar{\Delta}=-\Delta a^2/2$. The corresponding band diagonalizing matrix is $U= U({\bf k}, \bar{\Delta}) \oplus U({\bf k}, -\bar{\Delta})$, where 
\allowdisplaybreaks[4]
\begin{eqnarray}
U({\bf k}, \bar{\Delta})=\left( 
\begin{array}{cccc}
\frac{-\lambda_-}{\sqrt{2 \lambda \lambda_-}} & 0 & \frac{\lambda_+}{\sqrt{2 \lambda \lambda_+}} & 0 \\
\frac{v k_+}{\sqrt{2 \lambda \lambda_-}} & \frac{-\bar{g}({\bf k})}{\sqrt{2 \lambda \lambda_-}} & \frac{v k_+}{\sqrt{2 \lambda \lambda_+}} & \frac{-\bar{g}({\bf k})}{\sqrt{2 \lambda \lambda_+}} \\
0 & \frac{\lambda_-}{\sqrt{2 \lambda \lambda_-}} & 0 & \frac{-\lambda_+}{\sqrt{2 \lambda \lambda_+}} \\
\frac{\bar{g}({\bf k})}{\sqrt{2 \lambda \lambda_-}} & \frac{v k_-}{\sqrt{2 \lambda \lambda_-}} & \frac{\bar{g}({\bf k})}{\sqrt{2 \lambda \lambda_+}} & \frac{ v k_+}{\sqrt{2 \lambda \lambda_+}} 
\end{array} 
\right),
\end{eqnarray}
with $k_\pm = k_1 \pm i k_2$, $\lambda=\sqrt{v^2 k_+k_- + M^2 + \bar{g}({\bf k})^2}$, $\lambda_\pm =\lambda \pm M$ and $\bar{g}({\bf k})=\bar{\Delta}(k^2_1-k^2_2)$. In the presence of $\Delta_3$ pairing, the single-particle Hamiltonian in the vicinity of the Fermi surface for large mass ($M \gg v k, \Delta$) reads 
\allowdisplaybreaks[4]
\begin{eqnarray}
H^{\rm 2D}_{\rm FS} = \xi_{\bf k} \Gamma_{30} + \frac{\Delta_3}{M} \bigg\{ v \left[ k_x \Gamma_{11}-k_y \Gamma_{12} \right] 
+ \bar{g}({\bf k}) \Gamma_{20} \bigg\},
\end{eqnarray} 
where $\xi_{\bf k}=v^2 k^2/(2 M)- \mu$, and $\Gamma_{\rho \nu}=\eta_\rho \sigma_\nu$. The Pauli matrices $\{ \eta_\nu \} (\{ \sigma_\nu \})$ operate on the particle-hole (pseudospin) indices. For $\bar{\Delta}=0$, $H^{\rm 2D}_{\rm FS}$ describes a ${\mathcal T}$-symmetric fully gapped $p$-wave pairing. The $p$-wave BdG quasiparticles solely arises from the Dirac nature of NS fermions. If we set $\bar{g}({\bf k})=g_0$ (constant), then $H^{\rm 2D}_{\rm FS}$ describes a trivial $p+is$ pairing. The breaking of ${\mathcal T}$ and mixing of parity inside the paired state stem from the lack of ${\mathcal T}$ and ${\mathcal P}$ symmetries in the NS, respectively. By contrast, when $\bar{g}({\bf k})=\bar{\Delta}(k^2_1-k^2_2)$, $H^{\rm 2D}_{\rm FS}$ describes a HOT $p+id$ pairing, and the appearance of the $d$-wave component can now solely be attributed to the lack of $C_4$ symmetry in the NS. Therefore, the absence of each symmetry in the NS plays a crucial role in determining the symmetry and topology of the paired states~\cite{banddiagonalization-comment}. Note that $H^{\rm 2D}_{\rm FS}$ assumes the form of Dirac fermions ($p$-wave pairing), subject to an inverted-band regular Dirac mass (yielding a Fermi surface when $\mu>0$) and ${\mathcal P}$- and ${\mathcal T}$-odd, $C_4$ symmetry breaking WD mass ($d$-wave pairing); together giving rise to a HOTSC and four corner localized Majorana zero modes [Fig.~\ref{subfig1:delta3Delta}]. The same argument is applicable for the $\Delta_1$ pairing when $\theta=\frac{\pi}{2}$.

Similarly, the single-particle Hamiltonian for the three-dimensional PS pairing around the Fermi surface takes the form $H^{\rm 3D}_{\rm FS}=\xi_{\bf k} \Gamma_{30} + H^{\rm 3D}_{p+id}$, with $M=|m-3t_0|$ and  
\allowdisplaybreaks[4]
\begin{eqnarray}
H^{\rm 3D}_{p+id} = \frac{\Delta_{\rm ps}}{M} \bigg\{ v \left[ k_x \Gamma_{11}-k_y \Gamma_{12} + k_z \Gamma_{13}\right] 
+ \bar{g}({\bf k}) \Gamma_{20} \bigg\}.
\end{eqnarray} 
It takes the form of the ${\mathcal P}$- and ${\mathcal T}$-odd $p+id$ ($p+is$) pairing for $\bar{g}({\bf k})=\bar{\Delta}(k^2_1-k^2_2)$ (constant), standing as HOT (axionic~\cite{goswami-roy-axionSC}) superconductor. Therefore, around the Fermi surface the PS pairing is described by Dirac fermions ($p$-wave pairing) in the presence of both regular ($\xi_{\bf k}$) and ${\mathcal P}$- and ${\mathcal T}$-odd, $C_4$ symmetry breaking ($d$-wave pairing) Dirac masses, which gives birth to 1D hinge modes [Fig.~\ref{Fig:LDOS3D}]~\cite{banddiagonalization-comment}. Imprint of each discrete symmetry in the paired state is identical to that for the 2D HOTSC.

\emph{Discussion}. To summarize, here we demonstrate a general principle of realizing HOTSCs. In both $d=2$ and $3$, a first-order TSC can be converted into a HOTSC in the presence of a discrete symmetry breaking, ${\mathcal P}$- and ${\mathcal T}$-odd \emph{anticommuting} WD mass [Tables~\ref{Table:algebra_3D},~\ref{Table:algebra_2D}], respectively yielding corner and hinge modes of codimension $d_c=2$ [Figs.~\ref{Fig:LDOS2D1},~\ref{Fig:LDOS2D2},~\ref{Fig:LDOS3D}]. Therefore, magnetically doped or ordered (due to strong electronic correlations) Dirac insulators (trivial or topological) can accommodate HOTSCs, when they supports a Fermi surface (conducive for weak-coupling pairings) and are subject to external strain (breaking discrete rotational symmetry). Hopefully, the present discussion will stimulate a search for HOTSCs (both theoretically and experimentally) in strained but doped magnetic topological insulators~\cite{PTmass-latticecomment,moore-AFTI, regularDiracmasscomment}, such as MnBi$_2$Te$_4$~\cite{magnetic-TI-1}. In the future, it will also be worth investigating possible HOTSCs in candidate HOT insulators, such as Bi~\cite{schindler2018} and strained Luttinger materials~\cite{szabo}.

\emph{Acknowledgments}. B.R. was supported by the startup grant from Lehigh University.


\begin{thebibliography}{}
\bibitem{PTmass-latticecomment} Antiferromagnetism can lead to ${\mathcal P},{\mathcal T}$-odd Dirac mass.
\bibitem{comment-topology-NS} The topology of the NS plays no role in that for the paired states. All our conclusions remain unchanged when the NS is a topological insulator at zero doping.
\bibitem{banddiagonalization-comment} For the pairings that commute with the WD mass, the $p (p+id)$ component shows up in the intra-(inter)-band component. Consequently, their boundary modes, determined by the intra-band piece, are almost unaffected.
\bibitem{regularDiracmasscomment} Since the regular Dirac mass ($m$ or $M$) preserves all symmetries, it is generically finite in any Dirac insulator.
\end{thebibliography}
\end{document}